\documentstyle[12pt]{article}
\renewcommand{\theequation}{\thesection.\arabic{equation}}
\newlength{\extraspace}
\newlength{\extraspaces}
\newcommand{\be}{\begin{equation}
\addtolength{\abovedisplayskip}{\extraspaces}
\addtolength{\belowdisplayskip}{\extraspaces}
\addtolength{\abovedisplayshortskip}{\extraspace}
\addtolength{\belowdisplayshortskip}{\extraspace}}
\newcommand{\ee}{\end{equation}}
 
\newcommand{\ba}{\begin{eqnarray}
\addtolength{\abovedisplayskip}{\extraspaces}
\addtolength{\belowdisplayskip}{\extraspaces}
\addtolength{\abovedisplayshortskip}{\extraspace}
\addtolength{\belowdisplayshortskip}{\extraspace}}
\newcommand{\ea}{\end{eqnarray}}

\newcommand{\bas}{\begin{eqnarray*}
\addtolength{\abovedisplayskip}{\extraspaces}
\addtolength{\belowdisplayskip}{\extraspaces}
\addtolength{\abovedisplayshortskip}{\extraspace}
\addtolength{\belowdisplayshortskip}{\extraspace}}
\newcommand{\eas}{\end{eqnarray*}}
 
\newcounter{subequation}[equation]
\makeatletter

\expandafter\let\expandafter
\reset@font\csname reset@font\endcsname

\def\subeqnarray{\arraycolsep1pt
    \def\@eqnnum\stepcounter##1{\stepcounter{subequation}%
        {\reset@font\rm(\theequation\alph{subequation})}}
\jot5mm     \eqnarray}

\makeatother

\newcommand{\newsection}[1]{
\vspace{15mm}
\pagebreak[3]
\addtocounter{section}{1}
\setcounter{equation}{0}
\setcounter{subsection}{0}
\setcounter{footnote}{0}
 
\begin{flushleft}
{\large\bf \thesection. #1}
\end{flushleft}
\nopagebreak
\medskip
\nopagebreak}

\newcommand{\NP}[1]{Nucl.\ Phys.\ {\bf #1}}

\newcommand{\CMP}[1]{Comm.\ Math.\ Phys.\ {\bf #1}}

\newcommand{\C}{\mbox{$\,${\sf I}\hspace{-1.2ex}{\bf C}}}
\newcommand{\Z}{\mbox{{\sf Z}\hspace{-1ex}{\sf Z}}}
\newcommand{\R}{\mbox{\rm I\hspace{-.4ex}R}}
\newcommand{\1}{\mbox{1\hspace{-.6ex}1}}
\newcommand{\bra}{\langle}
\newcommand{\ket}{\rangle}
\newcommand{\ra}{\rightarrow}

\newcommand{\rra}{\ \longrightarrow \ }

\newcommand{\is}{ &\!=\!& }
\newcommand{\nonum}{\nonumber \\[1.5mm]}
\newcommand{\sspace}{\makebox[1cm]{ }}

\newcommand{\nspace}{\!\!\!\!\!\!\!\!\!\!}

\newcommand{\inv}{^{-1}}

\newcommand{\th}{{\theta}}
\newcommand{\lb}{\lambda}
\newcommand{\sh}{{\rm sh}}
\newcommand{\ch}{{\rm ch}}

\newcommand{\cA}{{\cal A}}

\newcommand{\cF}{{\cal F}}

\newcommand{\cH}{{\cal H}}

\newcommand{\cM}{{\cal M}}
\newcommand{\cO}{{\cal O}}

\newcommand{\cS}{{\cal S}}
 
\newcommand{\PHI}{{\widehat{\Phi}}}
\newcommand{\CH}{\underline{\cal H}}
\newcommand{\sX}{\scriptscriptstyle X}
\newcommand{\sY}{\scriptscriptstyle Y}

\newcommand{\sO}{\scriptscriptstyle O}

\begin{document}
%
\begin{titlepage}
%
\renewcommand{\thefootnote}{\fnsymbol{footnote}}
\begin{flushright}
MPI-PhT/97-31\\
\end{flushright}
\vspace{15mm}
 
\begin{center}
{\LARGE A Derivation of the Cyclic Form Factor Equation}\\[2.5cm]
{\large Max R. Niedermaier}\\ [3mm]
{\small\sl Max-Planck-Institut f\"{u}r Physik} \\
{\small\sl (Werner Heisenberg Institut)} \\
{\small\sl F\"{o}hringer Ring 6, D-80805 Munich, Germany}
\vspace{3.5cm}
 
{\bf Abstract}
\end{center}
\begin{quote}
A derivation of the cyclic form factor equation from quantum field 
theoretical principles is given; form factors being the matrix elements 
of a field operator between scattering states. The scattering states 
are constructed from Haag-Ruelle type interpolating fields with support 
in a `comoving' Rindler spacetime. The cyclic form factor equation then 
arises from the KMS property of the modular operators $\Delta$ 
associated with the field algebras of these Rindler wedges. 
The derivation in particular shows that the equation holds in any 
massive 1+1 dim. relativistic QFT, regardless of its integrability. 
\end{quote}
\vfill

\renewcommand{\thefootnote}{\arabic{footnote}}
\setcounter{footnote}{0}
\end{titlepage}
\newsection{Introduction}

Form factors of a $1+1$ dim. massive quantum field theory (QFT) and 
modular structures in the sense of algebraic QFT are apparently unrelated 
concepts. Form factors are matrix elements of some field operator between 
an asymptotic multi-particle state and the physical vacuum. As such
they are parametrized by a set of (initially real) rapidity variables 
$\th_j$ in which they admit a meromorphic continuation and possibly by
a set of internal quantum numbers $a_j,\;1\leq j\leq n$. We shall write 
$F_{a_n\ldots a_1}(\th_n, \ldots ,\th_1) =
\bra 0|\cO| \th_n,a_n;\ldots ;\th_1,a_1\ket$, where $\cO$ is a   
field operator (obtained by smearing a relativistic field,
possibly nonlocal and charged with a test function) and each rapidity 
$\th$ parametrizes
an on-shell momentum in the usual way, $p_0 = m\ch \th,\;p_1 = m\sh \th$. 
In theories with a factorized scattering operator there 
exists a system of functional equations for these form factors, which 
entail that the Wightman functions built from them have all the required 
properties, and which in principle allow one to compute the former exactly.  
Knowing the form factors, the Wightman functions can be reconstructed 
through convergent series expansions, which arise from inserting a 
resolution of the identity in terms of scattering states. Truncating 
the series at a finite particle number provides a powerful solution 
technique that produces non-perturbative results difficult or impossible 
to obtain otherwise. The most innocently looking of these functional 
equations is the cyclic form factor equation, stating that 
\be
F_{a_n\ldots a_1}(\th_n, \ldots ,\th_2,\th_1 +2\pi i) =
\eta\,F_{a_1a_n\ldots a_2}(\th_1,\th_n,\ldots,\th_2)\;,
\label{I1}
\ee
where $\eta$ is a phase and the shift by $2\pi i$ is understood in 
the sense of analytic continuation. Originally equation (1.1) was found 
in the context of the Sine-Gordon model
\cite{SmirSG} improving on earlier attempts to generalize Watsons 
equation \cite{KarWeisz}. Subsequently Smirnov promoted it to an axiom
for the form factors of an integrable QFT, which together with the 
other equations implies locality \cite{Smir}. The purpose of this paper 
is to give a derivation of equation (1.1) from quantum field theoretical 
principles. The derivation shows in particular that (1.1) holds in any 
massive 1+1 dim. relativistic QFT, regardless of its integrability.
The crucial tools are the modular structures (in the sense of
algebraic QFT) in a `Rindler wedge' situation, where they have geometrical
significance.

Modular structures in the context of von Neumann algebras are a 
pair of operators $(J,\Delta)$ that can be associated to any 
von Neumann algebra $\cM$ with cyclic and separating vector $\Omega$.
The latter means that there exists a Hilbert space $\cH$ 
such that both $\cM\Omega$ and $\cM'\Omega$ are dense subspaces 
of $\cH$, where $\cM'$ is the commutant of $\cM$. (The set 
and von Neumann algebra of all bounded operators on $\cH$ commuting 
with $\cM$.) The operator $J$ is an anti-unitary involution with 
respect to the inner 
product on $\cH$, and $\Delta$ is a positive selfadjoint (in general 
unbounded) operator. The defining relations for $(J,\Delta)$ 
are: $J \Delta^{1/2} X \Omega = X^* \Omega$ for $X \in \cM$ and 
$J \Delta^{-1/2} X' \Omega = {X'}^* \Omega$ for $X' \in \cM'$, where
$ J \Delta J = \Delta^{-1}$. The Tomita Takesaki theorem \cite{TT} states 
that $J \cM J = \cM'$ and that for all real $\lb$ the mapping
$D_{\lb}(X) = \Delta^{i\lb} X \Delta^{-i\lb}$ defines an automorphism 
group of both $\cM$ and $\cM'$. From the defining relations one
can deduce the following ``KMS property'' of $\Delta$ \cite{HHW,TT}
\be
(\Omega,Y\Delta X\Omega) = (\Omega, X Y\Omega)\;,\;\;\;X,Y \in \cM\;.
\label{I2}
\ee
Heuristically one can thus think of $\Delta$ as being an unbounded 
density operator for which the defining relations and (\ref{I2}) 
provide a substitute for the cyclic property of the trace.

In algebraic QFT one deals with a net of von Neumann 
algebras $\cM(K)$ associated to bounded regions $K$ (double cones) 
of the Minkowski spacetime and (by the Reeh-Schlieder theorem) for 
each $\cM(K)$ the vacuum provides a cyclic and separating vector. Hence the 
Tomita-Takesaki theory applies. The same holds when $K$ is replaced with a 
Rindler wedge, in which case the modular symmetries have geometrical
significance. Basically $J$ acts as a reflection and exchanges the 
left and the right Rindler wedge and $-\frac{1}{2\pi}\ln \Delta$ can 
be identified 
with the generator of Lorentz boosts along a direction that leaves 
the wedge invariant. Heuristically one can think of $\Delta$ as an
unbounded operator implementing Lorentz boosts with purely {\em imaginary} 
parameter and $J$ as being related to the CPT operator. In the framework 
of the Wightman QFT the above result is essentially due to Bisognano 
and Wichmann \cite{BW}, while in the more general algebraic setting an 
analogous 1+1 dim. result has more recently been proved by Borchers
\cite{Borch1}. 

In this context equation (\ref{I1}) is clearly reminiscent of the 
``KMS property'' (\ref{I2}) of the modular operator $\Delta$. For an 
actual derivation of (\ref{I1}) based on (\ref{I2}) one has to deal with 
three aspects of the problem. First one has to make sure that the 
action of the modular operator is defined on (vectors generated by) 
appropriate operators localized in a wedge domain and having sharply 
peaked momentum transfer. Second one has to show that these operators 
generate the usual scattering states of a Minkowski space QFT. 
Third, in order to cover reasonably generic QFTs, soliton sectors should be 
taken into account. This is because in 1+1 dim.~massive particles 
often have soliton properties, i.e.~interpolate between inequivalent 
vacua, and excluding them asymptotic completeness cannot be expected 
to hold; see e.g. \cite{Froeh}. 
It is the combination of these aspects which renders the derivation
of (\ref{I1}) technically a bit subtle. In the next section we describe 
the required general QFT framework in the presence of soliton sectors.
In section 3 we discuss some aspects of a Haag-Ruelle type scattering 
theory tailored towards the use of modular structures. 
The derivation proper of (\ref{I1}) is given in section 4.


\newsection{QFT framework including solitons}

Since equation (1.1) is a statement about matrix elements of 
scattering states, it is clear that the proper QFT framework for 
its derivation must ensure that the QFT under consideration has a 
well-behaved scattering theory. Apart from the set-up in which a 
Haag-Ruelle type scattering theory is formulated in higher dimensions, in
$1+1$ dim. this requires the inclusion of soliton sectors, because
otherwise asymptotic completeness cannot be expected to hold.
A model independent understanding of the appropriate QFT framework 
in the presence of soliton sectors was obtained only recently 
by Rehren and M\"{u}ger \cite{Mueger,Rehren} based on early 
work by Fr\"{o}hlich \cite{Froeh}. We shall adopt here 
a version of this framework suiting our purposes, the guideline
being more simplicity rather than minimality of the assumptions. 
The paragraphs containing major assumptions on the QFT considered 
are numbered (1) -- (6).

(1) The QFT is supposed to be described in terms of a net of local 
observables $K \ra \cA(K)$ satisfying isotony and locality \cite{Haag}. 
For simplicity we also require covariance with respect to the 
action of the proper $1+1$ dim. Poincar\'{e} group $P_+$. 
This means that there exists a representation 
of $P_+$ by automorphisms $p \ra \alpha_p$ such that 
$\alpha_p\cA(K)= \cA(p K)$. Elements $p =p(y,\lb,r)\in P_+$ 
can be parametrized by triples $(y,\lb,r)$, where $y \in \R^{1,1}$ is 
a translation parameter, $\lb \in \R$ is a boost parameter and 
$r \in \{\pm 1\}$ is a sign. Our conventions are 
$p(y,\lb,1)x = y + x(\lb)$, with 
$x^0(\lb) = x^0\ch \lb+ x^1\sh \lb$, $x^1(\lb) = x^0\sh\lb + x^1\ch\lb$,
and $p(0,0,\pm 1) x = \pm  x$. The subgroup generated by $p(y,\lb,1)$
is the restricted Poincar\'{e} group, denoted by $P_+^{\uparrow}$. 
For arguments and indices referring to  
$p \in P_+$ we will use the shorthands $y = p(y,0,1)$, $\lb = p(0,\lb,1)$
and $r=p(0,0,r)$. The $C^*$-algebra associated with a double cone $K$ of 
$1+1$ dim. Minkowski space is denoted by $\cA(K)$ and is assumed to be 
a factor. In $1+1$ dim. each double cone $K$ is an intersection of two 
translated wedges $K = (L + x) \cap (R + y)$, where 
$L= \{ x\in \R^{1,1} |\,|x^0| < - x^1\}$ and 
$R = \{ x\in \R^{1,1} |\,|x^0| < x^1\}$ are the left and right 
Rindler wedge. For an unbounded region $G$ let $\cA(G)$ denote 
the algebra obtained by taking the normclosure ($C^*$-inductive 
limit) of $\bigcup_{K\subset G} \cA(G)$; in particular 
$\cA = \cA(\R^{1,1})$ is the algebra of quasilocal observables. 
For any state $\omega$ over $\cA(G)$ we write 
$(\cH_{\omega},\pi_{\omega},\Omega)$ for the GNS triple 
of $\omega$ and denote by $\cM_{\omega}(G)$ the von Neumann algebra 
$\pi(\cA(G))''$, where the double prime denotes the 
weak closure in the $C^*$-algebra of bounded operators on $\cH_{\omega}$.
For the states $\omega$ of interest $\cH_{\omega}$ carries a 
positive energy representation of the Poincar\'{e} group. If 
$P_{\omega}$ denotes the generator of the translation subgroup in 
$\cH_{\omega}$ this means that its spectrum ${\rm Sp}(P_{\omega})$ 
is contained in the closed forward lightcone (spectrum condition).

(2) Specifically we assume that the QFT under consideration has both massive
1-particle and massive vacuum states. These concepts are defined as 
follows \cite{BF}. A massive 1-particle state is defined to a pure 
translation covariant state on $\cA$ such that the spectrum Sp$(P_{\omega})$ 
on $\cH_{\omega}$ consists of the mass shell $\{p|\,p_0 >0,\,p^2 = m^2\}$ 
and a subset of the continuum $\{p|\,p_0>0,\,p^2 = (m +\mu)^2\}$, for 
some $\mu >0$. Similarly a massive vacuum state is defined, except 
that Sp$(P_{\omega})$ now
consists of the value $0$ and a subset of $\{p|\,p_0>0,\,p^2 \geq\mu^2\}$,
where $\mu>0$ is called the mass gap. The unitary equivalence 
class of irreducible GNS representations associated with a given massive 
1-particle states is called a massive 1-particle sector and will be denoted 
by $a$ or $[\omega_a]$. The set of massive 1-particle sectors 
is assumed to be finite and is denoted by $I$. Similarly massive vacuum 
sectors $[\omega_{\alpha}]$ are defined, of which there may be 
infinitely many. For these vacuum sectors we shall assume that they obey 
wedge duality, i.e.
\be
\cM_{\alpha}(L+c) = \cM_{\alpha}(R+c)'\;,\;\;\;\forall c \in \R^{1,1}\;,
\ee 
where as usual the prime denotes the commutant in the algebra of 
bounded operators on the (separable) GNS Hilbert space. One can interpret 
$\cA(L+c)$ as a weakly dense subalgebra of $\cM_{\alpha}(L+c)$
and similarly for the right wedges. We do not require Haag-duality. 

If (2.1) is replaced with Haag duality, this is roughly also the 
$1+1$ dim. specialization of the set-up in which 
superselection sectors in $d+1$ dim. in the sense of DHR and BF
are discussed \cite{DHR,BF,Haag}. A peculiarity of $1+1$ dim. is that 
non-trivial superselection sectors in this sense do not exist \cite{Mueger} 
(under certain conditions which are supposed to be satisfied in 
massive QFTs). The massive 1-particle states can however have soliton 
character, i.e. interpolate between two inequivalent vacua at 
positive or negative spacelike infinity. This is related to a 
topological speciality of $1+1$ dim. Minkowski space: The spacelike 
complement of any double cone 
has two disconnected components, a left and a right component.
Associated with any massive 1-particle state $\omega_a$ are 
therefore a pair of massive vacuum states $\omega_{\alpha_L}$ and 
$\omega_{\alpha_R}$ \cite{Fproc}.

(3) Concerning the vacuum structure we assume that the different 
vacua arise (exclusively) from a spontaneously broken internal 
symmetry group $G$, as in \cite{Froeh}. For reasons that will 
become clear later we take the internal symmetry group to be abelian. 
More precisely $G$ is supposed to satisfy the following conditions:
(G1) Elements $g \in G$ are $*$-automorphims of $\cA$ preserving
the net structure, i.e. $g(\cA(K)) = \cA(K)$ for all double 
cones $K$. (G2) Elements $g \in G$ commute with the Poincar\'{e}   
group: $\alpha_p \circ g = g \circ \alpha_p$, $p \in P_+$.
(G3) $G$ is abelian, finitely generated and has trivial second
cohomology group. In concrete terms the latter means that every 
2-cocycle $e:G\times G \ra \C$ is a 2-coboundary, i.e.
\be
e(g_2,g_3) e(g_1g_2,g_3)^* e(g_1,g_2g_3) e(g_1,g_2)^* =1\,,
\ee
with $|e(g_1,g_2)|=1$ and  $e(g,\1) = 1$ implies 
$e(g,h) = \lambda(g)\lambda(h) \lambda(gh)^*$, for some 
1-cocycle $\lambda$ of $G$. Examples are the cyclic groups 
$\Z$ and $\Z_N$, $N>0$. (G4) The vacuum states
$\omega_{\alpha}$ and $\omega_{\alpha} \circ g$ are unitarily 
inequivalent for all $g \neq \1$. As indicated, it is convenient to
fix a reference vacuum state $\omega_{\alpha}$ and label all other 
vacua by group elements. In particular we shall write 
$(\cH_{\alpha\circ g},\pi_{\alpha \circ g},\Omega)$ for the 
GNS triple of $\omega_{\alpha}\circ g$. The representation of the 
Poincar\'{e} group is the same in all vacuum representations 
and is denoted by $p \ra U(p)$, with the shorthands $U(x) =
U(p(x,0,1))$, etc..

As described before massive 1-particle states will now interpolate 
between two such vacua at left and right spacelike infinity.
In many situations one will be interested only in the interpolation 
properties of a state, not in its particle properties. This motivates 
to define kink states as follows: A state $\omega$ over $\cA$ is called 
a kink state, interpolating between vacuum states $\omega_{\alpha} \circ g$ 
and $\omega_{\alpha} \circ h$, if it is a translation covariant state 
satisfying the spectrum condition and if it has the property
\be
\pi_{\omega}\Big|_{\cA(L)} \sim \pi_{\alpha\circ g}\Big|_{\cA(L)}\;,\sspace
\pi_{\omega}\Big|_{\cA(R)} \sim \pi_{\alpha\circ h}\Big|_{\cA(R)}\;,
\ee
where `$\sim$' denotes unitary equivalence. Naturally a kink
sector is an equivalence class $[\omega]$ of kink states. Note that
massive 1-particle states are special kink states. Following Fr\"ohlich 
\cite{Froeh} we next assume that (all) kink states can be constructed 
from vacuum states by means of suitable automorphisms of $\cA$ 
whose existence is postulated.

(4) For any pair $(g,h) \in G \times G$ we assume that there exists
a $*$-automorphism $\rho$   (a ``kink automorphism of 
type $(g,h)$'') enjoying the following properties:
\begin{itemize}
\item[($\rho 1$)] There exists a bounded double cone $K$ such that
\be
\rho\Big|_{\cA(K_L)} = g\Big|_{\cA(K_L)}\sspace {\rm and}\sspace 
\rho\Big|_{\cA(K_R)} = h\Big|_{\cA(K_R)}\;,
\label{out1}
\ee
where $K_L$ and $K_R$ are the left and right spacelike complement 
of $K$, respectively. The region $K$ is called the interpolation 
region of $\rho$.   
\item[($\rho 2$)] $\rho$ commutes with the symmetries: $\rho g = g \rho$.
\item[($\rho 3$)] There exists a strongly contineous map
$\gamma_{\rho}:P^{\uparrow}_+ \ra \cA$ (a cocycle) such that  
\begin{subeqnarray}
&& \gamma_{\rho}(p) A \gamma_{\rho}(p)^{-1} =
(\alpha_p\circ \rho\circ \alpha_{p\inv} \circ \rho^{-1})(A),\;\;
A \in \cA\;,\\
&& \gamma_{\rho}(p_2p_1) = 
\alpha_{p_2}(\gamma_{\rho}(p_1))\gamma_{\rho}(p_2)\;,\\
&& g \gamma_{\rho}(p) = \gamma_{\rho}(p)\;,\sspace
\gamma_g(p)=\1\;, \;\;\; g \in G\;.
\label{out2}
\end{subeqnarray}
\item[($\rho 4$)] There exists a group homomorphism 
$G\times G \ni (\1,h) \ra \rho$, denoted by $h \ra \rho_h$.
\end{itemize}

Let us add a few comments. The absence of non-trivial DHR sectors 
\cite{Mueger} implies that $\rho$ is determined by its type 
$(g,h)$ up to unitary equivalence. That is to say, if $\rho_1,\rho_2$ 
are two automorphisms of type $(g,h)$ then $\rho_1 \rho_2^{-1}$ 
is an inner automorphism of $\cA$. Condition ($\rho 4)$ thus says that
out of each unitary equivalence class one can pick a representative
such that $h \ra \rho_h$ becomes a group homomorphism. Then 
${}_g\rho_h := \rho_{h g^{-1}}g$ is a (preferred) kink automorphism 
of type $(g,h)$. The supplementary 
condition (\ref{out2}c) on the cocycles is only included for convenience; it 
could be relaxed and then follows from the other properties. Concerning 
($\rho 1$) one verifies that $\gamma_{\rho}(p)$, $p=p(\lb,x,1)$  has 
interpolation region $x+p(\lb,0,1) K$, if $K$ is the interpolation region 
of $\rho$. Further one checks that the set of kink automorphisms forms 
a group with  respect to composition. In particular the inverse of a 
kink automorphism of type $(g,h)$ is of type $(g^{-1}, h^{-1})$ and 
has the same interpolation region.
Parallel to the DHR case \cite{DHR} one can show that two kink
automorphisms commute if their interpolation regions are spacelike 
separated. Clearly a necessary condition for this to happen is that
the group $G$ is abelian, which supplemented by ($\rho 2$) also 
turns out to be sufficient \cite{Rehren}. 
If the interpolation regions of $\rho_1$ and $\rho_2$ are not 
spacelike separated, $\rho_1 \rho_2$ and $\rho_2 \rho_1$ are related by a 
unitary ``statistics operator'' as in the DHR case,
$\rho_1 \rho_2 = {\rm Ad}\epsilon(\rho_1,\rho_2)\circ \rho_2 \rho_1$.
The latter is defined by separating the interpolation regions by means 
of unitary, gauge invariant charge transporters; by (\ref{out2}c) the 
cocycles serve that purpose. It then follows that  
$\epsilon(\rho_1,\rho_2)$ depends at most on the orientation of the 
auxiliary spacelike separated regions employed in the separation process. 
In particular it satisfies $\epsilon({}_{g_1}\rho_{h_1}, 
{}_{g_2}\rho_{h_2}) = \epsilon(\rho_{h_1g_1^{-1}},\rho_{h_2g_2^{-1}})$,
and turns out always to be a complex phase. 
The statistics phase proper is defined as $\kappa_{\rho} = 
\epsilon(\rho,\rho)$ and obeys $\kappa_{\rho} = \kappa_{\rho^{-1}}=
\kappa_{\rho \circ g}$, $\forall g \in G$. From $(\rho 4)$, the trivial 
second cohomology of $G$, and the properties of the statistics operator 
one can show that there exists a choice of square roots 
$\sqrt{\kappa_{\rho}}$ such that 
\be
\epsilon(\rho_1,\rho_2) = \frac{\sqrt{\kappa_{\rho_1\rho_2}}}%
{\sqrt{\kappa_{\rho_1}}\sqrt{\kappa_{\rho_2}} }\;.
\label{phase1}
\ee

The main use of the kink automorphisms lies in the fact
that they generate kink states from vacuum states. In detail, let 
$\rho$ be a kink automorphism of type $(g,h)$ and let $\omega_{\alpha}$ be 
a massive vacuum state. Then the state $\omega_{\alpha}\circ \rho$ 
is a kink state interpolating the vacuum sectors 
$[\omega_{\alpha}\circ g]$ and $[\omega_{\alpha}\circ h]$ $(*)$.
Let us briefly comment on the proof of $(*)$. The fact that 
the state $\omega_{\alpha}\circ {}_g\rho_h$ has the correct interpolation 
property is manifest. Its translation covariance follows from the 
translation part of the identity
\ba
&& (\pi_{\alpha}\circ \rho \circ \alpha_p)(A) = 
U_{\rho}(p)\,(\pi_{\alpha}\circ \rho)(A)\,U_{\rho}(p)^{-1}\;,\sspace
{\rm where}\nonum
&& U_{\rho}(p) := U(p) \pi_{\alpha}(\gamma_{\rho}(p\inv))\;.
\label{out3}
\ea
Here $p \ra U_{\rho}(p)$ is the representation of the (restricted) 
Poincar\'{e} group in $\pi_{\alpha \circ \rho}:= \pi_{\alpha} \circ \rho$,
which can thus be constructed from $U$ and the cocycle. 
To complete the proof of $(*)$ it remains to establish the spectrum 
condition, which is done in \cite{Froeh,Schlingel}.

Using the kink automorphisms therefore all superselection sectors, 
that is all of the relevant representations of $\cA$ can be realized
on a fixed reference Hilbert space, as in the DHR case \cite{DHR,BF}.
In detail, pick a reference vacuum state and let $(\cH_{\alpha},\pi_{\alpha},
\Omega)$ denote its GNS triple. For an interpolating automorphism $\rho$
consider the representation $\pi_{\alpha} \circ \rho$ with Hilbert space
$\cH_{\alpha \circ\rho}$. An automorphism $\hat{\rho} = {\rm Ad}V \circ \rho$
unitarily related to $\rho$ induces an unitarily equivalent 
representation $\pi\circ \hat{\rho} = {\rm Ad}\, W \circ (\pi \circ \rho)$,
$W = \pi(V)$ and vice versa. The space of cone-localized unitary intertwiners 
$V: \rho \ra \hat{\rho}$ is denoted by $(\hat{\rho}|\rho)$. A 
generalized state is an equivalence class of pairs 
\be 
(\pi, \psi_{\alpha}) \sim ({\rm Ad}W\circ \pi, W\psi_{\alpha})\;,\;\;\;
W=\pi(V),\;V \in (\hat{\rho}|\rho) \;, 
\label{gstate}
\ee
where $\psi_{\alpha} \in \cH_{\alpha}$ and $\pi = \pi_{\alpha} \circ
\rho$ for some interpolating automorphism $\rho$. 
We shall use $[\pi,\psi_{\alpha}]$ to denote the equivalence class 
generated by the pair  $(\pi,\psi_{\alpha})$. Each equivalence class
$[\pi,\psi_{\alpha}]$ defines a state over $\cA$ by means of the assignment
\be
\cA \ni A \rra \frac{(\psi_{\alpha}, \pi(A)\psi_{\alpha})}%
{(\psi_{\alpha},\psi_{\alpha})}\;.
\label{gstate2}
\ee
The classes ${\bf \Omega}_{\alpha \circ g}:=[\pi_{\alpha}\circ g, \Omega]$ 
play the role of the vacua. The inner product of two pairs  
$(\pi, \psi)$ and $(\pi',\psi')$ is declared to vanish when 
$\pi$ and $\pi'$ are not unitarily equivalent. Otherwise it is defined 
to be $(\psi,\psi')$ for representatives such that $\pi = \pi'$. 
In particular the norm 
of $(\pi,\psi)$ is the norm of $\psi$. The space of all pairs 
$(\pi,\psi_{\alpha})$ equipped with this inner product and 
norm is called the state bundle and is denoted by $\CH$.  

The assumption ($\rho 4)$ allows one to choose a global section 
in this bundle. To see this let $\rho_h, h \in G$ be the collection  
of automorphisms forming a representation of $G$. As noted before, 
then ${}_g\rho_h := \rho_{hg^{-1}} g$ is an automorphism of 
type $(g,h)$ and any other of the same type is unitarily equivalent 
to it. Letting now  $(g,h)$ run through $G \times G$, each sector 
is visited once and only once. If we denote by 
${}_g\!\cH_h$ the Hilbert space of pairs $({}_g\rho_h, \psi_{\alpha})$,
$\psi_{\alpha} \in \cH_{\alpha}$, the direct sum $\bigoplus_{g,h} 
{}_g\!\cH_h$ provides a global section through the state bundle 
${\underline \cH}$.

Next we define an extension of the observable algebra acting
irreducibly on ${\underline \cH}$. It is obtained from pairs 
$(\rho,A)$ consisting of a kink automorphism and a quasilocal
operator $A \in \cA$. They act on pairs $(\pi,\psi_{\alpha})$ 
by $(\rho, A) (\pi, \psi_{\alpha}) = (\pi \circ \rho, 
\pi(A)\psi_{\alpha})$. The associated generalized state 
$[(\rho, A) \circ (\pi, \psi_{\alpha})]$ defines a kink state over
$\cA$ by (\ref{gstate2}). The equivalence relation 
(\ref{gstate}) induces a corresponding one 
$({\rm Ad}V \circ \rho, V A) \sim (\rho,A)$ 
on the pairs $(\rho,A)$. The set of 
such pairs can be given the structure of an associative 
$*$-algebra with multiplication and $*$-operation given by 
\ba
&& (\rho_2,A_2) (\rho_1,A_1) =
(\rho_1\rho_2,\rho_1(A_2)A_1)\;,\nonum
&& (\rho, A)^* = (\rho^{-1}, \rho^{-1}(\Phi^*))\sim
(\bar{\rho}, \bar{\rho}(A^*)V_r)\;. 
\label{kinkalg1}
\ea
The automorphism $\bar{\rho}$ entering in the second line is 
defined by $\bar{\rho} = \alpha_r \circ \rho \circ \alpha_r \circ 
(gh)^{-1}$, where $\rho$ is of type $(g,h)$. Further $\alpha_r$ is
the automorphism of $\cA$ associated with $r=p(0,0,1)\in P_+$ and 
$V_r$ is a unitarity. $\bar{\rho}$ is of type $(g^{-1}, h^{-1})$
and has interpolation region $-K$, if $K$ is the interpolation region 
of $\rho$. The former implies that $\bar{\rho}$ is unitarily 
related to $\rho^{-1}$, i.e. $\bar{\rho} = {\rm Ad}V_r \circ \rho^{-1}$,
but has reflected interpolation region. In either version, the 
$*$-operation is compatible with the inner product on ${\underline \cH}$. 
We shall refer to this algebra as ``kink algebra'' and 
denote it by $\cF$. It can be given a net structure satisfying
Poincar\'{e} covariance and isotony. The action of the 
(restricted) Poincar\'{e} group is  
\be
(\rho,A) \stackrel{p}{\rra} (\rho, A_{\rho}(p)) =: 
{\bf U}(p)(\rho,A) {\bf U}(p)^{-1}\;,\;\;\;{\rm with}\;\;\; 
A_{\rho}(p) := \gamma_{\rho}(p)^* \alpha_p(A)\;,
\label{out4}
\ee
where $\gamma_{\rho}$ is a $P_+^{\uparrow}$-cocycle for $\rho$. 
As anticipated by the notation,  the action (\ref{out4})
of the Poincar\'{e} group on pairs $(\rho,A)$ commutes with the 
composition law and the $*$-operation (\ref{kinkalg1}) 
due to the cocycle identity (\ref{out2}a). Concerning the 
localization properties, we say that a kink operator $(\rho,A)$ 
is localized in a double cone $K$ if there exists a representative 
$(\hat{\rho}, \widehat{A})$ in the unitarity equivalence class
$({\rm Ad}V \circ \rho, V A) \sim (\rho, A)$   
such that $\hat{\rho}$ has interpolation region 
$K$ and $\widehat{A}\in \cA(K)$. With this definition one shows that
$(\rho,A_{\rho}(x))$ is localized in $x +K$, if 
$(\rho,A)$ is localized in $K$, and that the product
of two kink operators is localized in the smallest double cone
containing the localization regions of the individual operators.
The $*$-algebra of kink operators localized in $K$
is denoted by $\cF(K)$, the subspace where $\rho$ is of type $(g,h)$
by ${}_g\cF_h(K)$.  

Further the kink algebra carries two (mutually commuting) internal group
actions. First, the original spontaneously broken symmetry, which is 
implemented unitarily on $\cF$ via
\be
(\rho,\Phi) \stackrel{g}{\rra} (\rho, g(\Phi)) = 
{\bf Q}_g^{-1} (\rho, \Phi){\bf Q}_g =: g(\rho,\Phi)\;,
\label{kinkalg2}
\ee
with ${\bf Q}_g =(g,\1)$. In particular ${\bf Q}_g$ connects the 
different vacuum sectors in ${\underline \cH}$ by 
${\bf Q}_g{\bf \Omega}_{\alpha} ={\bf \Omega}_{\alpha \circ g}$.
Second, there is an unbroken dual symmetry acting on $\cF$ by 
$({}_g\rho_h, \Phi) \ra \chi(g,h)({}_g\rho_h, \Phi)$, where
$\chi(g,h)$ is a character of $G\times G$. Both symmetries 
preserve the localization and commute with the $*$-operation and 
Poincar\'{e} transformations.     

So far we have concentrated on the interpolation properties
of the elements of the kink algebra $\cF$. We now select those 
elements of $\cF$ that generate interesting 1-particle states
from a vacuum ${\bf \Omega}_{\alpha \circ g}$. First this requires 
$\rho$ to be such that $\omega_{\alpha} \circ g\rho$ is a 
massive 1-particle state. In addition $\Phi \in \cA$ must be chosen such 
that the spectral support of this state is contained in the mass 
shell $\{p\,|\,p_0>0,\, p^2 = m_a^2\}$. Finally it is natural to assume 
that $\Phi$ transforms irreducibly both under the action of the 
Lorentz group and under the action of $G$. Whence
\be
{\bf U}(\lb) {\bf A} {\bf U}(\lb)^{-1} = e^{s_a\lb}\,{\bf A}\;,
\sspace  
{\bf Q}_g^{-1} {\bf A} {\bf Q}_g= \chi_a(g)\,{\bf A}\;,
\label{kinkalg3}
\ee
where $s_a \in \R$ is called the spin of ${\bf A}$ and $\chi_a \in 
\widehat{G}$ the character of ${\bf A}$. As indicated we reserve
an extra symbol ${\bf A} =(\rho, \Phi)$ for these elements of $\cF$
and call them soliton operators, or 1-kink operators, of type 
$a=(g,h;m_a,s_a,\chi_a)$.
(The special case where $\rho$ actually interpolates between 
equivalent vacua, i.e. where ${\bf A}$ isn't a soliton operator
proper, is included in this terminology.) 
The set of 1-kink operators ${\bf A}$ of type $a$ with 
interpolation region $K$ is denoted by $\cF_a(K)$. 
By construction a soliton operator generates a 1-particle 
state from a vacuum sector.  The 1-particle subspace of 
${\underline \cH}$ of type $a$ is denoted by $\cH_a^{(1)}$. 
For the soliton operators one computes the following 
exchange relations \cite{Rehren}
\ba
&& {\bf A}\, {\bf B} = \delta_{ab}(\pm)\,{\bf B}\, {\bf A}\;,\sspace
\pm(K_a -K_b)\succ 0\;,\nonum
&& \delta_{ab}(+) 
=\epsilon(\rho_a,\rho_b)\, \chi_b^*(g_a)\chi_a(h_b)\;,\nonum
&& \delta_{ab}(-) 
=\epsilon(\rho_b,\rho_a)^*\,\chi_b^*(h_a)\chi_a(g_b)\;.
\label{kinkalg4}
\ea 
Here `$\succ$' denotes the partial ordering for double cones, i.e.
$\widetilde{K} \succ K\;\Leftrightarrow\;\widetilde{K}-K\subset R$.  
The relations (\ref{kinkalg3}) and (\ref{kinkalg4}) clearly generalize
to kink operators that are arbitrary products of soliton operators
and which carry the induced quantum numbers.

(5) For the discussion of Spin-Statistics and the construction of a 
CPT operation it seems indispensable at present to assume that the 
kink algebra $\cF$ is generated by non-local Wightman fields in the 
following sense: There are limits of operators $(\rho, A) \in \cF(K)$ 
of pointlike lokalization, which upon translation give rise to 
non-local Wightman fields ${\bf F}(x)$, being unbounded operator-valued
distributions. These fields then inherit the algebraic structures 
on $\cF$ (group actions, multiplication and $*$-operation, interpolation
properties, exchange relations etc.). In particular to 
each Wightman field ${\bf F}(x)$ a kink automorphism $\rho$ (with 
pointlike interpolation region) is associated and, after decomposition 
into irreducible components, also a group character $\chi$ and a 
Lorentz spin $s$. Cone-localized operators ${\bf F}$ can be recovered 
by averaging with appropriate test functions. We retain the previous 
terminology by saying that ${\bf F}$ is localized in $K$, if ${\bf F}$ 
arises from averaging a field ${\bf F}(x)$ with a test function supported 
in $K$ and if the kink automorphism associated with ${\bf F}$ can be 
chosen to have interpolation region $K$. In a slight abuse of notation 
we shall also write $K \ra \cF(K)$ for the local net (of unbounded 
operators) generated thereby, and continue to call the elements kink 
operators. In principle one can always switch to bounded counterparts 
of these operators and we assume that the net obtained thereby coincides 
with the previously defined kink algebra of bounded operators.
For the purposes here the distinction between both descriptions is 
only essential for the construction of a CPT operation.
In preparation of the latter, let ${\bf F}_a(x)$, ${\bf F}_b(x)$ be 
two (Wightman) soliton fields of type $a,b$, respectively. Their 
two-point function obeys \cite{Rehren}
\ba
&& \left({\bf F}_a(x){\bf \Omega}_{\alpha}\,,
\,{\bf F}_b(y){\bf \Omega}_{\alpha}\right) = \omega_a \omega_b^* \,
\left({\bf F}_b(-y)^*{\bf \Omega}_{\alpha}\,,
\,{\bf F}_a(-x)^*{\bf \Omega}_{\alpha}\right)\;,
\;\;\; (x -y)^2 <0\;,\nonum
&&{\rm where}\sspace  \omega_a =  \frac{e^{i\pi s}}{\sqrt{\kappa_{\rho}} }\,  
\chi(h)^* =  e^{-i\pi s} \sqrt{\kappa_{\rho}}\, \chi(g)^*\;. 
\label{kinkalg5}
\ea
Both expressions for $\omega_a$ coincide by the following spin-statistics 
relation \cite{Rehren}
\be
e^{2\pi i s} = \kappa_{\rho}\, \chi(hg^{-1})\;.
\label{kinkalg6}
\ee
This is to say, the spin of a soliton field of type $(g,h;m,s,\chi)$ is 
determined up to an integer by $gh^{-1}$ and the character $\chi$.
In order to get non-vanishing matrix 
elements in (\ref{kinkalg5}) $\rho_a$ and $\rho_b$ have 
to be of the same type $(g,h)$, in which case the phase in
(\ref{kinkalg5}) only depends on the common unitary equivalence class
$\rho \sim {\rm Ad}V \circ \rho$. In contrast, the character is not 
super selected, i.e. $\chi_a\neq \chi_b$, $\chi_a,\chi_b \in \widehat{G}$
does not enforce the matrix element (\ref{kinkalg5}) to vanish. If the 
character were super selected only ``neutral'' operators of trivial
character could have a non-vanishing vacuum expectation value and 
consequently the vacuum states $\omega_{\alpha \circ g}$ were 
all equal. From (\ref{kinkalg5}) one can anticipate the proper 
definition of the CPT operation, which for expositional reasons
we defer to section 4.

(6) Finally we assume `completeness of the particle picture' 
in the following sense. First, there are sufficiently many soliton 
operators/fields  in $\cF_a(K)$ such that upon averaging the translated 
operators with rapidly decaying wave functions all of $\cH_a^{(1)}$ can be 
generated from a vacuum sector. Second we assume the following version 
of asymptotic completeness. Given the collection of 1-particle Hilbert 
spaces $\cH_a^{(1)},\;a\in I$ one can apply a standard second quantization 
procedure to them, resulting in a Fock space. For the purposes here it 
is convenient to work with the free (`unsymmetrized') 
Fock space $F$. Off hand the Fock space $F$ is completely unrelated to 
the physical Hilbert space $\CH$. The Haag-Ruelle theory in this context 
provides a constructive way to isometrically embed two distinguished 
(proper or improper) subspaces $\CH^{\rm ex},\;{\rm ex} ={\rm in/out}$ 
of $\CH$ into $F$. We assume that the image in $F$ can be identified 
with subspaces of `rapidity ordered' wave functions 
$F^{\rm ex},\;{\rm ex} ={\rm in/out}$; c.f. section 3, step 3.
Since $F^{\rm in}$ and $F^{\rm out}$ are isometric this entails 
asymptotic completeness, i.e. $\CH^{\rm in} =\CH^{\rm out}$.


\newsection{Aspects of a Haag-Ruelle Scattering theory in $1+1$ dim.}

Here we describe those aspects of a Haag-Ruelle type scattering 
theory in $1+1$ dimensions required for the derivation in section 4. 
Compared to $3 +1$ dimensions there are two technical complications.
First, the convergence for $t \ra \pm \infty$ of the states built
from the multi-particle interpolating fields is only guaranteed for
velocity ordered configurations. Second, the particle concept itself
is more complicated due to the existence of solitons. As remarked before,
the inclusion of soliton states is crucial for the discussion 
of scattering theory, because otherwise asymptotic completeness cannot
be expected to hold. The assumption (5) is not needed here and the 
DHR description is used throughout this section.

The construction basically involves three steps:
\begin{enumerate}
\item Construction of 1-particle interpolating fields. 
\item Construction of multi-particle scattering states.
\item Verification that the norms of these states factorize, \newline
yielding isometric embeddings $\CH^{\rm ex}\ra F$. 
\end{enumerate}

Tailored towards the use of geometric modular structures we wish 
to use ingredients localized in a wedge domain, which requires an 
approximation procedure. In the following we describe the 
so-adapted steps 1. -- 3. consecutively.

1. Construction of the 1-particle interpolating fields: 
Let $(\rho,\Phi)\in \cF_a(K)$ be a soliton operator of type $a$ and let
$\PHI_{\rho}(p) := \int d^2 x \,e^{-ip\cdot x} \Phi_{\rho}(x)$ be the Fourier 
transform of the translated operator $\Phi_{\rho}(x)$. We 
define the 1-particle interpolating field by    
\ba
&& {\bf A}(f^t|\th) = \left(\rho, A(f^t|\th) \right)\;,\sspace
A(f^t|\th)=\int\frac{d^2p}{(2\pi)^2}\,\widehat{f}^t(p)\,
\PHI_{\rho}(p)\;,\;\;\;\;{\rm where}\nonum
&& \widehat{f}^t(p) = \widehat{f}(p)\;e^{i(p_0 -\omega(p_1))t}\;,
\sspace\omega(p_1)=\sqrt{p_1^2 +m_a^2}. 
\label{A1}
\ea
Here $\widehat{f}(p)$ is a energy-momentum distribution with 
the following features: It is smooth (infinitely differentiable) with 
compact support in $\R^{1,1}$ and non-vanishing connected intersection
with the mass hyperboloid $p^2 = m_a^2$. For $\delta >0$ we define the 
velocity support of $f$ by 
\be 
v_{\delta}(f) =
\{ v(p) := p_1/\omega(p_1)\,|\, \|p -k\| \leq \delta ,\;k \in 
{\rm supp}(\widehat{f}) \},
\label{vsupp}
\ee
where $v(p)$ is the velocity with respect to the Lorentz frame
determined by the $x^0$-coordinate. In 1+1 dim. it is convenient to 
use coordinates $p_0 =\mu \cosh\th ,\;p_1 =\mu\sinh\th$ on the forward 
lightcone, in which case the velocity is parametrized by the rapidity
$v(p) = \tanh \th$. In particular $v_{\delta}(f)$ determines 
some closed rapidity interval. We shall refer to the center of this 
rapidity interval as the ``average rapidity'' $\widehat{\th}$.
We also find it convenient to split the information contained in 
$\widehat{f}(p)$ into two parts: First an equivalence 
class of translated functions $\th \ra 
\widehat{f}(\mu\cosh(\th-\lb),\mu\sinh(\th-\lb))$ for some $\lb \in \R$;
and second the average rapidity $\widehat{\th}$ of 
$\widehat{f}(p)$, which determines a unique member of this equivalence 
class. In the notation $A(f^t|\widehat{\th})$ adopted in (\ref{A1}), the 
first argument refers to the equivalence class and the second to the 
average rapidity. The advantage of this notation is that Lorentz boosts 
act on the fields (\ref{A1}) basically by shifting the average 
rapidity, i.e.
\be
\gamma_{\rho}(\lb)^*\alpha_{\lb}\big(A(f^t|\th)\big)= 
e^{s_a\lb}\,A(f^t|\th +\lb)\;.
\label{A2}
\ee

Let us now address the localization properties of the 1-particle 
interpolating field ${\bf A}(f^t|\th)$. In position space the 
expression (\ref{A1}) for $A(f^t|\th)$ becomes
\ba 
&& A(f^t|\th) = \int d^2 x\, f^t(x) \Phi_{\rho}(x)\;,\;\;\;
{\rm where}\nonum
&& f^t(x) = \int\frac{d^2p}{(2\pi)^2}\,\widehat{f}^t(p)\,
e^{-i p\cdot x} = \int d^2 y D^t(x-y) f(-y)\;,\nonum
&& D^t(x) =  \int\frac{d^2p}{(2\pi)^2}\,e^{i(p_0 -\omega(p_1))t}\,
e^{-i p\cdot x}\;. 
\label{A3}
\ea 
Here $f$ is the Fourier transform of $\widehat{f}$ (but for notational
simplicity $f^t$ is the Fourier transform of $\widehat{f}^t$ with sign
reversed arguments). Since $\widehat{f}$ has compact support in momentum
space, $f$ and $f^t$ will not have compact support in position 
space, but will only be of `fast decrease'. In particular $A(f^t|\th)$
is only a quasilocal field, not an element of any algebra $\cA(K)$
associated with a bounded double cone $K$. With hindsight to the 
application of modular operators in a Rindler wedge situation we wish 
to approximate $A(f^t|\th)$ by local fields. In preparation let us 
examine the decay properties of $f^t(x^0,x^1)$ in more detail. A standard 
integration by parts argument shows that it  
decays faster than any power of $|t -x^0|^{-1}$ for $|t -x^0| \ra \infty$
with $x^1$ fixed. Similarly, for fixed $t$ it decays faster than any 
inverse power of $x^1$ for $x^1 \ra \infty$. Of particular interest is 
the limit along trajectories of the form $x^0 = t,\;x^1 = -v t$, with 
$v \not\in v_{\delta}(f)$. Ruelle's lemma \cite{HR} states that $f^t(t,-vt)$ 
decays faster than any inverse power of $t$ for $|t| \ra \infty$.%
\footnote{A quick check on the signs is via the stationary phase
approximation.} This motivates to introduce compact regions
\be
G^{t,\delta}(f) = \{ x \in \R^{1,1} \,|\, x^0 \in [ t-\delta, t+\delta],
\; x^1 \in - x^0\, v_{\delta}(f) \}\;,
\label{xsupp}
\ee
whose spatial extension grows linearly in $|t|$. For a soliton 
operator $(\rho,\Phi)\in \cF_a(K)$ then define
\be
{\bf A}^{\delta}(f^t|\th) = \left(\rho,A^{\delta}(f^t|\th) \right)\;,
\sspace A^{\delta}(f^t|\th) = \int_{G^{t,\delta}(f)} d^2 x 
\,f^t(x) \Phi_{\rho}(x)\; 
\label{A4}  
\ee
and the bounded double cone
\be
K^{t,\delta} = {\rm cone}\left( 
\bigcup_{x \in G^{t,\delta}(f)} (x + K)\right)\;,
\label{icone}
\ee
where ${\rm cone}(G)$ denotes the smallest double cone containing the set 
$G$. One can then show: 

(a) ${\bf A}^{\delta}(f^t|\th)$ has interpolation
region $K^{t,\delta}$. (b) The norm of the difference of the fields
(\ref{A3}) and (\ref{A4}) is bounded by some rapidly decaying function 
$d(t)$, i.e. $\|A(f^t|\th) - A^{\delta}(f^t|\th)\| < d(t)$. 

The proof of (a) can be found in \cite{Schlingel}; we only add 
that the use of Haag duality can be avoided, consistent with our
assumptions in section 2. In a slight abuse of notation we shall temporarily
use $(\rho, A^{\delta}(f^t|\th))$ also to denote the representative 
$({\rm Ad}V^t\circ \rho, V^t A^{\delta}(f^t|\th))$ ($V^t$ a 
cone-localized unitarity) for which the 
automorphism has interpolation region $K^{t,\delta}$ and the operator is 
an element of $\cA(K^{t,\delta})$. Given Ruelle's lemma in the form 
\be
\int_{R^{1,1} \backslash G^{t,\delta}(f)} d^2 x \, f^t(x) < d(t)\;,
\label{bound0}
\ee
the proof of (b) amounts to  
\ba
&& \|A(f^t|\th) - A^{\delta}(f^t|\th)\| = 
\bigg\|\int_{R^{1,1}\backslash G^{t,\delta}(f)} d^2 x\, 
f^t(x) \Phi_{\rho}(x)\bigg\| \nonum 
&& \leq \|\Phi\| \, \int_{R^{1,1}\backslash G^{t,\delta}(f)} 
d^2 x \,f^t(x) < d(t) \|\Phi\|\;.
\label{bound1}
\ea

2. Construction of multi-particle scattering states: 
Let $(\rho_j,\Phi_j) \in \cF_{a_j}(K_j)$, $1\leq j\leq n$, be a collection
of soliton operators with interpolation regions $K_j$ to be specified 
later. Let ${\bf A}_j(f^t|\th) = \left(\rho_j, A_j(f^t|\th) \right)$ 
be the associated 1-particle interpolating fields and 
${\bf A}_j^{\delta}(f^t|\th) = \left(\rho_j, 
A_j^{\delta}(f^t|\th) \right)$ be the approximants (\ref{A4}). 
Using the composition law (\ref{kinkalg1}) the product fields can be 
computed, for which we introduce the shorthands 
\ba
{\bf X}^{t,\delta} \is {\bf A}_n^{\delta}(f^t_n|\th_n) \ldots 
{\bf A}_1^{\delta}(f^t_1|\th_1) =:(\rho_1\ldots \rho_n,\,X^{t,\delta})\;,\nonum
{\bf X}^t \is {\bf A}_n(f^t_n|\th_n) \ldots {\bf A}_1(f^t_1|\th_1)=:
(\rho_1\ldots \rho_n,\,X^t)\;,
\label{short3}
\ea
for the restricted and unrestricted case, respectively. 
For the reference vacuum ${\bf \Omega}_{\alpha} = 
[\pi_{\alpha}, \Omega]$ consider the states 
${\bf X}^t {\bf \Omega}_{\alpha}$ and ${\bf X}^{t,\delta} 
{\bf \Omega}_{\alpha}$. We wish to arrange the data on
which these states depend such that for $t\ra \infty$ they 
converge in norm to states in the physical Hilbert space $\CH$. The   
norm of pairs $(\rho,A)$ or $(\pi,\Psi)$ here is simply defined 
as the norm of the second entry of the pair. The convergence can be 
achieved by an appropriate choice of the localization regions $K_j$ of 
the 1-particle operators and the velocity supports $v_{\delta}(f_j)$ 
of the wave functions. The proper requirements are 
\begin{subeqnarray}
&& K_n \prec K_{n-1} \prec \ldots \prec K_1\;,\\
&& v_n < v_{n-1} < \ldots < v_1\;, \sspace \forall v_j \in 
v_{\delta}(f_j)\;.
\label{cond1}
\end{subeqnarray}
The states ${\bf X}^{t}{\bf \Omega}_{\alpha}$ with data (\ref{cond1}) are 
the 1+1 dim.~version of Hepp-Ruelle ``non-overlapping states''. For the 
restricted fields ${\bf X}^{t,\delta}$ the condition (\ref{cond1}b) 
guarantees that the ordering (\ref{cond1}a) translates into an ordering 
of the bounded interpolation regions (\ref{icone})
\be 
K_n^{t,\delta} \prec K_{n-1}^{t,\delta} \prec \ldots \prec K_1^{t,\delta}\;,
\label{cond2}
\ee
for large enough $t>0$. Further the spatial distance between these
double cones tends to infinity as $t\ra \infty$. On the other hand  
one has the multi-particle generalization of (\ref{bound1})
\be
\|{\bf X}^{t,\delta}{\bf \Omega}_{\alpha} - {\bf X}^t{\bf \Omega}_{\alpha}\| 
<d(t)\;.
\label{bound3}
\ee
Combining (\ref{cond2}) and (\ref{bound3}) one can follow 
the classic arguments \cite{HR,DHR} to show that 
\be
\bigg\| \frac{d}{dt} {\bf X}^t{\bf \Omega}_{\alpha} \bigg\| < d(t)\;,
\label{bound2}
\ee
for some rapidly decreasing function $d(t)$. A more detailed account can
be found in section 6.3 of \cite{Schlingel}. From (\ref{bound2})
one concludes that the family of vectors ${\bf X}^t{\bf \Omega}_{\alpha}$
converges strongly for $t\ra \infty$ to a vector in $\CH$, which is the
searched for candidate for an  $n$-particle `out' scattering state. 
It turns out to depend only on the 1-particle input data; in particular
it is easily checked to be independent of the choice of the Lorentz frame 
used. Further, by (\ref{bound3}) the restricted interpolating fields 
generate the same scattering states. To adhere to the rapidity notation 
usually employed in the context of form factors, we shall describe the 
limits in terms of improper momentum eigenstates as follows 
\be
{\bf \Psi}^{\rm out}:= 
\lim_{t \ra \infty} {\bf X}^{t,\delta}{\bf \Omega}_{\alpha} =
\lim_{t \ra \infty} {\bf X}^t{\bf \Omega}_{\alpha}  
=: \int \frac{d^n\th}{(4\pi)^n}\,f_{n}(\th_n)\ldots f_{1}(\th_1)\,
|\th_n,a_n;\ldots, \th_1,a_1\ket ^{\rm out} \;, 
\label{Sstates}
\ee 
where $f_j(\th)$ stands for $f_j(m_{a_j}\cosh\th,m_{a_j}\sinh\th)$, 
$j=1,\ldots n$,  and the massive 1-particle representations are $\pi_{a_j}=
\pi_{\alpha} \circ \rho_j$. For simplicity we treat only `out' scattering 
states here. For `in' scattering states some of the ordering relations
have to be reversed. Since we assume asymptotic completeness, it is 
convenient to treat them  as CPT transforms of the `out' states, as 
we shall do later.

3. Isometric embedding $\CH^{\rm out}\ra F$: It remains to show 
that the norm of the limiting vectors (\ref{Sstates}) factorizes
into a product of terms depending only on the 1-particle input
data. As usual this follows from clustering, since the conditions
(\ref{cond1}) ensure that the spatial distances of the essential support
regions (\ref{xsupp}) tend to infinity as $|t|\ra \infty$ \cite{HR,DHR,BF}. 
Details in the case at hand can be found in \cite{Schlingel}.
This factorization entails that the limiting states (\ref{Sstates}) 
can be identified with certain Fock space vectors having the same norm. 
In the momentum space description used before an $n$-particle vector is 
represented by a wave function $f^{(n)}(\th_n, a_n;\ldots ;\th_1,a_1)$ 
with ordered and separated rapidities $\th_n < \ldots < \th_1$, together 
with an  assignment $(a_n, \ldots , a_1)$ to particle types. 
The space of sequences of such functions forms a subspace of the 
free Fock space $F$ (where no relations among the creation and 
annihilation operators are imposed) built from the 1-particle Hilbert 
spaces. Explicitly
\ba
&& F^{\rm out}\subset F = \bigoplus_{n=1}^{\infty} \bigoplus_{I^n}\,
\cH^{(1)}_{a_n} \otimes \ldots \otimes \cH^{(1)}_{a_1}\;,\nonum
&& F^{\rm out} =\Big\{ (f^{(n)})_{n\geq 0}\,|\,{\rm supp}\,f^{(n)}\in 
\{\th \in \R^n |\th_n < \ldots < \th_1\}\Big\}\;.
\label{Fock1}
\ea 
The inner product on $F$ is inherited from the 1-particle sectors. 
The inner product on 1-particle states of type $a,b$ is 
\ba
&&(A(f_1|\th_1)\Omega,\,A(f_2|\th_2)\Omega)=
\delta_{a,b}\int_{-\infty}^{\infty}\frac{d\th}{4\pi}\, 
f^*_1(\th) f_2(\th)\;\;\Longleftrightarrow\;\;\nonum 
&& {}^{\rm out}\!\bra\th_1,a|\th_2,b\ket^{\rm out} = 4\pi 
\delta_{a b} \delta(\th_1 -\th_2)\;.
\label{Fock0}
\ea
The isometric embedding $\CH^{\rm out}\ra F$ obtained thereby is a 
somewhat weaker result as in the usual Haag-Ruelle theory. The reason
is that in $\CH^{\rm out}$ additional relations among the state 
vectors exist, which result from the exchange relations (\ref{kinkalg4}) 
of the field operators (in coordinate space) used 
in their construction. Correspondingly the image $F^{\rm sym}$ of 
$\CH^{\rm out}$ in $F$ induced by (\ref{Sstates}) will consist of 
sequences of wave functions obeying certain `symmetry' relations. 
In momentum space their explicit description may be cumbersome. 
Nevertheless one expects that these relations allow one to extend 
the domain of definition of an $n$-particle momentum space wave 
function from, say, the sector $\{\th\in\R^n | \th_n < \ldots <
\th_1 \}$ to all of $\R^n$, while preserving the norm. 
This is what the assumption in paragraph (6) of section 2 amounts to. 
In other words, we can view the exchange relations (\ref{kinkalg4}) as 
defining an isometry between $F^{\rm sym}$ and $F^{\rm out}$. For the 
purposes here only the final isometry between $\CH^{\rm out}$ and 
$F^{\rm out}$  matters.

\newsection{Cyclic form factor equation and modular structures}

After these lengthy preparations we now turn to the derivation proper 
of (1.1). The idea is to use the modular operators of a family 
of (right) wedge domains $R^t = c^t + R$ shifted along a path
$t \ra c^t \in \R^{1,1}$, such that the restricted interpolating 
fields at time $t$ have support in $R^t=  c^t + R$ and the action of 
geometric modular operators is defined. In this way equation (\ref{I1}) 
arises from the ``KMS property" (\ref{I2}) of the modular operator 
$\Delta$. As a guideline let us recall how (\ref{I2}) arises from the 
defining relations of $(J,\Delta)$. The latter are: 
$J \Delta^{1/2} X \Omega = X^* \Omega$ for $X \in \cM$ and 
$J \Delta^{-1/2} X' \Omega = {X'}^* \Omega$ for $X' \in \cM'$, where
$ J \Delta J = \Delta^{-1}$. From this and the anti-unitarity of 
$J$ one obtains (\ref{I2}) via \cite{HHW,TT} 
\ba
&& (\Omega, Y \Delta X\Omega) = (\Delta^{1/2} Y^*\Omega\,,
\Delta^{1/2} X \Omega) \nonum 
&& =(J Y \Omega\,,J X^*\Omega) =
(X^*\Omega\,,Y\Omega) = (\Omega, X Y\Omega)\;,\;\;\;X,Y \in \cM\,.
\label{KMS1}
\ea
The aim in the following is to transfer this computation to 
the situation at hand. To this end one first has to ensure that
the $n$-particle interpolating fields have support in $R^t$ and that 
the action of $\Delta^{1/2}$ is defined on the vectors generated by 
them. 
\smallskip

Let ${\bf A}_j^{\delta}(f_j^t|\th_j)$, $1\leq j\leq n$ be a collection
of 1-particle interpolating fields with data satisfying (\ref{cond1}). 
Let $(\rho_j,A_j^{\delta}(f_j^t|\th_j))$ be the representatives 
for which $\rho_j$ has interpolation region $K_j^{t,\delta}$ and 
$A_j^{\delta}(f_j^t|\th_j)\in \cA(K_j^{t,\delta})$. For 
a suitably chosen $z^t\in \R^{1,1}$ we shall be interested in the 
analyticity properties in $\lb$ of vectors of the form 
$U_n(\lb)U_n(-z^t)\pi_{\alpha}(X^{t,\delta})\Omega$,
with $X^{t,\delta}$ as in (\ref{short3}). Writing this vector out 
explicitly, the kink representations $\pi_j := \pi_{\alpha} \circ 
\rho_1\ldots \rho_j$ appear and we will use $U_j$ to denote 
$U_{\rho_1\ldots\rho_j}$ in (\ref{out3}). One finds 
\ba
&\nspace & U_n(\lb)U_n(-z^t)\pi_{\alpha}(X^{t,\delta})
\Omega\nonum
&& = e^{(s_{a_n}+\ldots + s_{a_1})\lb}\, 
\int_{G_n^t}d^2 y_n \ldots  \int_{G_1^t}d^2 y_1 \;
f^t_n(y_n)\ldots  f^t_1(y_1)\;U_n(y_n(\lb) - z^t(\lb))\times \nonum
&&\times 
\pi_{n-1}(\Phi_n) U_{n-1}(y_{n-1}(\lb) - y_n(\lb))\ldots 
\pi_1(\Phi_2)U_1(y_1(\lb) - y_2(\lb))\pi_{\alpha}(\Phi_1) \Omega\;.
\label{vector}
\ea 
Here $z^t\in \R^{1,1}$ is chosen such that $-z^t + G_n^t \subset R$ and to 
simplify the notation we wrote $G_j^t$ for $G^{t,\delta}(f_j)$, 
$1\leq j\leq n$. The guideline to determine the analyticity properties  
of (\ref{vector}) is the following simple fact. Let $p\ra U(p)$ be a strongly
continous unitary representation of $P^{\uparrow}_+$ on a separable 
Hilbert space obeying the spectrum condition. Consider $U(x(\lb)) =
U(\lb)U(x)U(\lb)^{-1}$ for $x\in \R^{1,1}$, with the notation
$x^0(\lb) = x^0\ch \lb+ x^1\sh \lb$, $x^1(\lb) = x^0\sh\lb + x^1\ch\lb$. 
Then 
\be
\lb \ra U(x(\lb)) \;\;\mbox{is analytic in}\;\;
\left\{ \begin{array}{ll}
\phantom{-}0<{\rm Im}\,\lb< \pi\;,\;\; &{\rm if}\; x \in R\;,\\ 
-\pi<{\rm Im}\,\lb< 0 \;,\;\; & {\rm if}\; x \in L\;.
\end{array}
\right.
\label{strips}
\ee
Further $U(x(\lb))$ is a bounded operator in these strips. 
Applied to the vector (\ref{vector}) one sees that the dependence on 
$\lb$ is analytic in the strip $0 <{\rm Im}\lb < \pi$. 
Indeed, since the spatial distance between the regions $G^t_j$ 
increases with $t$, there exists a $t_0 > 0$ such that 
${\rm cone}(G_n^t) \prec \ldots \prec {\rm cone}(G_1^t)$, for
$t \geq t_0$, which implies $y_j - y_{j+1} \in R$ for $j =1, \ldots, n-1$ 
with $y_j \in G_j^t$. For the argument of $U_n$ the condition 
$y_n - z^t \in R$ holds by definition of $z^t$. 

In summary, we found that for a suitably chosen $z^t \in \R^{1,1}$
the support regions $G_j^t = G^{t,\delta}(f_j)$ are contained 
in a shifted right wedge domain $z^t + R$, for $t > t_0$. 
In this shifted wedge $U_{n,z^t}(\lb) := 
U_n(z^t) U_n(\lb)U_n(-z^t)$ plays the role of the Lorentz boost generator
and acts consistently on the vector $\pi_{\alpha}(X^{t,\delta})
\Omega$ for $0 \leq {\rm Im}\,\lb \leq \pi$. The localization 
regions $K_j^{t,\delta}$ of the 1-particle interpolating fields are not 
necessarily contained in $z^t +R$. However, since they are 
likewise compact regions, related to $G_j^t$ by (\ref{icone}),
one can find a $c^t \in \R^{1,1}$ (timelike and 
future-pointing) and $t_1 \geq t_0$ such that $K_j^{t,\delta} 
\subset c^t + R$, for all $t > t_1$. With this definition of 
$R^t := c^t + R$ one has $A_j^{\delta}(f_j^t|\th_j) \in 
\cA(K_j^t)\subset \cA(R^t)$. It is easy to see that such localization 
properties are preserved under composition of kink operators.
For the multiparticle interpolating fields (\ref{short3}) one can thus 
choose representatives $\rho_{\sX} ={\rm Ad}V^t\rho_1\ldots \rho_n$ 
($V^t$ a cone-localized unitarity) having interpolation region 
${\rm cone}(K_1^{t,\delta} \cup \ldots 
\cup K_n^{t,\delta})$ and $X(R^t) := V^t X^{t,\delta}\in \cA(R^t)$.
Having ensured that such a choice of representatives is possible,
the outcome of the previous discussion is conveniently 
recast in terms of kink operators and generalized states.  
Set 
\be
{\bf U}_t(\lb) := {\bf U}(c^t) {\bf U}(\lb) {\bf U}(-c^t)\;,
\;\;\;{\bf \Delta}_R^s := {\bf U}_t(2\pi i s)\;,\;\;\;
{\bf \Delta}_L^s := {\bf U}_t(-2\pi i s)\;,\;\;s >0\;.
\ee

{\bf Proposition 1} 

(a) Let ${\bf A}_j^{\delta}(f_j^t|\th_j)$, $1\leq j\leq n$, be 
restricted 1-particle interpolating fields with data satisfying (\ref{cond1}).
Then there exist wedge domains $R^t = c^t + R$ and $t_1 > 0$ 
such that the restricted  $n$-particle interpolating field
\be 
{\bf X}(R^t) := {\bf A}_n^{\delta}(f^t_n|\th_n) \ldots 
{\bf A}_1^{\delta}(f^t_1|\th_1) =
\left(\rho_{\sX}, X(R^t)\right)\;,  
\label{wedgefields}
\ee
has bounded interpolation region in $R^t$ for all $t > t_1$.
Symbolically ${\bf X}(R^t) \in \cF(R^t)$. 

(b) ${\bf \Delta}_R^s$ is a positive densely defined 
operator on $\cF(R^t){\bf \Omega}_{\alpha}$, for all
$0 \leq s \leq 1/2$.   
 
(c) The generalized states ${\bf X}(R^t){\bf \Omega}_{\alpha}$
converge strongly to scattering states in $\CH$ for $t \ra \infty$. 
\smallskip

Until here the assumption (5) did not enter. Now we employ it to 
construct a CPT operation on $\cF$ in its Wightman version. 
Recall the notation ${\bf Q}_g =(g,\1)$ and let ${\bf Q}$ be 
the unitary involution on $\CH$, acting like ${\bf Q}_{(gh)^{-1}}$
on the sector ${}_g\!\cH_h$. Define an operator ${\bf \Theta}$ on 
$\CH$ and  ${\rm Ad}{\bf \Theta}$ on $\cF$ by  
\be 
{\bf \Theta} {\bf F}(x) {\bf \Omega}_{\alpha} = \omega_a {\bf Q} 
\,{\bf F}(-x)^* {\bf \Omega}_{\alpha}\;,\sspace 
{\bf \Theta} {\bf F}(x) {\bf \Theta} = \omega_a {\bf Q}_{gh} 
{\bf F}(-x)^*\;,
\label{CPT}
\ee
where in the second equation ${\bf F}(x)$ is of type $(g,h)$
and hence ${\bf Q}_{gh} {\bf F}(-x)^*$ is of type $(h,g)$.
Then ${\bf \Theta}$ has the following properties:
\smallskip

{\bf Proposition 2} 

(a) ${\rm Ad}{\bf \Theta}$ is an antilinear $*$-automorphism of 
$\cF$ and an involution. Further ${\bf \Theta}$ is antiunitary
w.r.t. the inner product on $\CH$.

(b) The following commutation relations hold
\be
\begin{array}{rclcrcl}
{\bf \Theta} {\bf U}(\pm i\pi) \is {\bf U}(\mp i\pi) {\bf \Theta}\;,
& \sspace &
{\bf \Theta} {\bf Q}_g         \is {\bf Q}_g {\bf \Theta}\;, 
\\
{\bf \Theta} {\bf U}(\lb)      \is {\bf U}(\lb) {\bf \Theta}\;,
& \sspace &
{\bf \Theta} {\bf U}(x)        \is {\bf U}(-x) {\bf \Theta}\;.
\end{array}
\ee

(c) ${\bf \Theta} \cF(R) {\bf \Theta} = \cF(L)$ and vice versa.
\smallskip 

The proof can be adapted from Rehren \cite{Rehren}.
This CPT operator arises, up to a unitary factor, from the polar 
decomposition of the following Tomita operators
\ba 
&& {\bf S}_+ {\bf F} {\bf \Omega}_{\alpha} = {\bf Q}\, h({\bf F}^*) 
{\bf \Omega}_{\alpha}\;, \sspace {\bf F} \in {}_g\cF_h(R) \;,\nonum
&& {\bf S}_- {\bf F} {\bf \Omega}_{\alpha} = {\bf Q}\, g({\bf F}^*) 
{\bf \Omega}_{\alpha}\;, \sspace {\bf F} \in {}_g\cF_h(L) \;.
\label{Tom1}
\ea
In fact, the closures of the operators (\ref{Tom1}) can be seen to 
have adjoints related by $({\bf S}_{\pm})^* = \kappa^{\pm 1}{\bf S}_{\mp}$
and to admit polar decompositions
\be 
{\bf S}_{\pm} = {\bf J}_{\pm} {\bf U}(\pm i\pi)\;,\;\;\;
{\rm with} \;\;\; {\bf J}_{\pm} = \sqrt{\kappa}^{\pm 1} \,
{\bf \Theta}\;.
\label{Tom2}
\ee 
Here the unitary operator $\sqrt{\kappa}$ is declared to act by 
multiplication with $\sqrt{\kappa_{\rho}}$ on $\cH_{\alpha \circ \rho}$.
The origin of the unitary factor $\sqrt{\kappa}^{\pm 1}$ 
can be understood from the relations 
\ba
&& {\bf \Theta} {\bf U}(i\pi) {\bf F} {\bf \Omega}_{\alpha} = 
\sqrt{\kappa}^{-1} {\bf Q}\, h({\bf F}^*) {\bf \Omega}_{\alpha}\;,
\sspace {\bf F} \in {}_g\cF_h(R)\;,\nonum
&& {\bf \Theta} {\bf U}(-i\pi) {\bf F} {\bf \Omega}_{\alpha} = 
\sqrt{\kappa}\, {\bf Q}\, g({\bf F}^*) {\bf \Omega}_{\alpha}\;,
\sspace {\bf F} \in {}_g\cF_h(L)\;.
\label{Tom3}
\ea

Next we show that the CPT operation declared via 
(\ref{CPT}) on the kink operators induces a  CPT operation 
on scattering states having all the required properties.
The CPT conjugate of a 1-particle (Wightman) interpolating field 
of type $(g,h)$  naturally is
\ba
&& {\bf \Theta}\left(\rho,A^{\delta}(f^t|\th)\right){\bf \Theta} = 
\left( j(\rho), A^{\delta,CPT}({f^*}^{-t}|\bar{\th})\right)\;,
\sspace{\rm where} \nonum
&& j(\rho) = \bar{\rho} gh\;,\;\;\;  
A^{\delta}_{\rm CPT}({f^*}^{-t}|\bar{\th}) = \omega_a
\int_{G^{-t,\delta}(f^*)} d^2 y \,{f^*}^{-t}(y)\Phi^*(y)\;.
\label{CPT2}
\ea
We have displayed the representatives localized in 
$-K^{t,\delta} = -G^{t,\delta}$ and rewrote the operator such that 
the time reversal is manifest. The complex conjugate $f^*$ of $f$ 
plays the role of the charge 
conjugate wave function, whose average rapidity is denoted by $\bar{\th}$. 
Of course the spacetime reflection here is with respect to the 
origin of the chosen coordinate system and exchanges $R$ with $L$,
rather than the `comoving' wedge domains $R^t=c^t +R$ and 
$c^t +L =: L^t$. A CPT operation doing the latter is 
\be 
{\bf \Theta}_t := {\bf U}(c^t) {\bf \Theta} {\bf U}(-c^t)\;.
\ee
Let then ${\bf X}(R^t)$ be an $n$-particle interpolating field as 
in (\ref{wedgefields}) and consider its CPT conjugate 
\be
{\bf \Theta}_t{\bf X}(R^t) {\bf \Theta}_t = 
\left( j(\rho_n) \ldots j(\rho_1),\,
X(R^t)_{\rm CPT}\right) =: {\bf X}(R^t)_{\rm CPT}\;.
\ee
One easily sees that there exist representatives, displayed in the 
middle term, for which $X(R^t)_{\rm CPT} \in \cA(L^t)$ and 
$j(\rho_1)\ldots j(\rho_n)$ has bounded interpolation region in 
$L^t := c^t + L$ for $t < - t_1$. As before one can use them to study 
the analyticity properties in $\lb$ of the Lorentz boosted state 
${\bf U}(\lb){\bf U}(-c^t)\,{\bf X}(R^t)_{\rm CPT}{\bf \Omega}_{\alpha}$ 
as in (\ref{vector}). With the data for ${\bf X}(R^t)$ as in 
proposition 1, the dependence on $\lb$ is found to be analytic in the strip
$-\pi <{\rm Im}\lb < 0$. It follows that the action of 
${\bf \Delta}_L^s$, $0\leq s\leq 1/2$ is defined on 
${\bf X}(R^t)_{\rm CPT}$. Since the latter generate $\cF(L^t)$ 
one concludes that 
\be
{\bf \Theta}_t\, \cF(R^t) {\bf \Theta}_t = \cF(L^t) \;,\sspace
{\bf \Theta}_t {\bf \Delta}_L^s {\bf \Theta}_t = {\bf \Delta}_R^s \,,\;\;
0\leq s \leq 1/2\,,\;\;{\rm on}\;\;\cF(R^t) {\bf \Omega}_{\alpha}\;,
\ee
Further
\ba
&& {\bf S}_R {\bf X} {\bf \Omega}_{\alpha} = {\bf Q}\,h({\bf X}^*)
{\bf \Omega}_{\alpha}\;,\sspace {\bf X}\in 
{}_g\cF_h(R^t)\;,\;\;\;{\rm with}\;\;
{\bf S}_R = \sqrt{\kappa}{\bf \Theta}_t {\bf \Delta}_R^{1/2}\;,\nonum
&& {\bf S}_L {\bf X} {\bf \Omega}_{\alpha} = {\bf Q}\,g({\bf X}^*)
{\bf \Omega}_{\alpha}\;,\sspace {\bf X}\in 
{}_g\cF_h(L^t)\;,\;\;\;\;{\rm with}\;\;
{\bf S}_L = \sqrt{\kappa}^{-1}{\bf \Theta}_t {\bf \Delta}_L^{1/2}\,.
\ea
In particular the state ${\bf X}(R^t)_{\rm CPT}{\bf \Omega}_{\alpha}$
converges strongly to a scattering state in $\CH$ for $t\ra \infty$. 
On the improper scattering states (\ref{Sstates}) the following 
CPT operation is induced 
\be
{\bf J} |\th_n,a_n;\ldots ;\th_1,a_1\ket^{\rm out} 
= |\bar{\th}_1,\bar{a}_1;\ldots ;\bar{\th}_n,\bar{a}_n\ket^{\rm in} 
= |\bar{\th}_n,j(a_n);\ldots ;\bar{\th}_1,j(a_1)\ket^{\rm in}\;.
\label{Jscatt}
\ee
Here $a_k$, $\bar{a}_k$ and $j(a_k)$ refer to the massive 1-particle 
representations $\pi_{\alpha}\circ \rho_k$, $\pi_{\alpha}\circ
\bar{\rho}_k$ and $\pi_{\alpha}\circ j(\rho_k)$, respectively. 
Further $\th$ and $\bar{\th}$ are the average rapidities of a 
momentum space wave function $\widehat{f}$ and its complex conjugate, 
respectively.   
From (\ref{Jscatt}) one readily checks that ${\bf J}$ has all the familiar 
properties of a CPT operation on scattering states. In particular it 
leaves the scattering operator invariant ${\bf J \cS J} = {\bf \cS}^{-1}$ 
and the scattering operator ${\bf \cS}$ itself can be written as a product 
of ${\bf J}$ and the free CPT operator on the Fock space; see also
\cite{Schroer}.

Having all these ingredients at our disposal we can eventually 
transfer the computation (\ref{KMS1}) to the case at hand. 
Introduce generalized operators ${\bf X}(R^t),\;{\bf Y}(R^t)$ by
\ba 
&& {\bf X}(R^t):= {\bf A}_{n}^{\delta}(f^t_{n}|\th_{n}) \ldots 
{\bf A}_{n-k+1}^{\delta}(f^t_{n-k+1}|\th_{n-k+1}) =
\left(\rho_{\sX},X(R^t)\right)\;,\nonum
&& {\bf Y}(R^t):= {\bf A}_{n-k}^{\delta}(f^t_{n-k}|\th_{n-k}) \ldots 
{\bf A}_{1}^{\delta}(f^t_{1}|\th_{1}) =
\left(\rho_{\sY},Y(R^t)\right)\;,
\label{XY1}
\ea
where the data $f_j$ and $K_j$, $1\leq j\leq n$ are as in proposition 1
and the terms on the right denote the representatives with 
interpolation region $K_{\sX} := {\rm cone}(K_n^{t,\delta} \cup \ldots
\cup K_{n-k +1}^{t,\delta})$ and $K_{\sY} := 
{\rm cone}(K_{n-k}^{t,\delta} \cup \ldots \cup K_1^{t,\delta})$,
respectively. Further let ${\bf O} = (\rho_{\sO},\cO)$ be a   
kink operator and choose $d^t \in \R^{1,1}$ such that the translated 
operator ${\bf O}(d^t) := {\bf U}(d^t) {\bf O} {\bf U}(-d^t)$ has 
interpolation region $K_{\sO}$ satisfying 
$K_{\sX} \prec K_{\sO} \prec K_{\sY}$, for large $t$. 
In order to have nonvanishing matrix elements of the form required,
these operators have to satisfy an appropriate `charge balance' condition.
Explicitly, for some $k \in G$ we assume that 
\be
\rho_{\sY} \rho_{\sO} \rho_{\sX} = k\;,
\label{balance}
\ee
and write ${\bf \Omega}_{\beta} := {\bf \Omega}_{\alpha \circ k}$.
Further we abbreviate momentarily ${\bf X}= {\bf X}(R^t)$, 
${\bf Y}= {\bf Y}(R^t)$. Using (\ref{balance}) one verifies that all 
the matrix elements in the following chain of equalities are 
well-defined 
\ba
&\nspace\!& \left( {\bf \Delta}_R^{1/2}\,{\bf Y}^*
{\bf O}^*(d^t) {\bf \Omega}_{\beta}\,,\,
{\bf \Delta}_R^{1/2}\,{\bf X}\,{\bf \Omega}_{\alpha} \right) = 
\left( {\bf \Theta}_t\sqrt{\kappa}^{-1}{\bf S}_R\,
{\bf Y}^*{\bf O}^*(d^t)\,{\bf \Omega}_{\beta}\,,\,
{\bf \Theta}_t \sqrt{\kappa}^{-1} 
{\bf S}_R\,{\bf X}{\bf \Omega}_{\alpha}\right)
\nonum
&\nspace\!&  = 
\left({\bf S}_R {\bf X}\,{\bf \Omega}_{\alpha}\,,\,
{\bf S}_R {\bf Y}^*\,{\bf O}^*(d^t) {\bf \Omega}_{\beta} \right)
 = \left(k^{-1}h_{\sX}({\bf X}^*)\,
{\bf \Omega}_{\beta}\,,\,
h_{\sY}^{-1}h_{\sO}^{-1}({\bf O}(d^t){\bf Y})\,
{\bf \Omega}_{\alpha} \right)\,.
\label{XY4}
\ea
In the last expression we extract the character phases using 
(\ref{kinkalg3}) and then exchange the order of ${\bf X}$ and 
${\bf O}(d^t)$ using (\ref{kinkalg4}). Reinserting into 
(\ref{XY4}) results in the following identity
\be
\eta \left({\bf O}^*(d^t){\bf X}^*\,{\bf \Omega}_{\beta}\,,\,
{\bf Y}\,{\bf \Omega}_{\alpha} \right) 
= \eta \left({\bf \Omega}_{\beta}\,,\,
{\bf O}(d^t){\bf X}{\bf Y}{\bf \Omega}_{\alpha}\right)
= \left( {\bf Y}^*{\bf O}^*(d^t) {\bf \Omega}_{\beta}\,,\,
{\bf U}_t(2\pi i)\,{\bf X}\,{\bf \Omega}_{\alpha} \right) \;, 
\label{XY5}
\ee
where $\eta= \chi_{\sX\sO\sY}(h_{\sX}k^{-1})\,\delta_{\sX\sO}(-)$
is the accumulated phase, depending both on the statistics 
phases and the group characters of the involved kink operators.
The first expression in (\ref{XY5}) 
in particular shows that the $t\ra \infty$ limit of 
these matrix elements exists and yields well-defined matrix elements 
between scattering states. Since ${\bf O}$ is a cone-localized operator 
each of the matrix elements is separately well-defined 
also for $d^t =0$. On the other hand $\eta$ depends (for given 
kink operators) only on the orientation of the interpolating 
automorphisms and in particular is independent of $d^t$. The identity 
(\ref{XY5}) thus remains valid when sending $d^t$ to zero. 
Adopting the notation from (\ref{Sstates}), 
(\ref{A2}) and (\ref{Jscatt}) one arrives at
\ba
&\nspace\nspace\nspace &\eta\, 
{}^{\rm in}\!\bra \bar{\th}_{n-k+1} -i\pi,\bar{a}_{n-k+1};\ldots ;
\bar{\th}_n -i\pi,\bar{a}_n
|\, \cO\, |\th_{n-k},a_{n-k}; \ldots ;\th_{1},a_{1}\ket^{\rm out} 
\nonum &\nspace\nspace& 
\;\;\; = \eta\, {}^{\rm out}\!\bra 0|\, \cO\,
|\th_n,a_n;\ldots ; \th_1,a_1\ket^{\rm out} 
\nonum &\nspace\nspace&
\;\;\; = {}^{\rm out}\!\bra 0|\,\cO\,|
\th_{n-k},a_{n-k};\ldots;\th_{1},a_{1};\;
\th_n+i2\pi,a_n; \ldots ;\th_{n-k+1}+i2\pi ,a_{n-k+1}\ket^{\rm out} 
\label{ffcycl}
\ea
for ordered and separated rapidities, i.e. $\th_j-\th_{j+1}>\epsilon,\;
j=1,\ldots, n-1$ with some positive constant $\epsilon$. Both the 
``crossing relation'' and the ``cyclic form factor equation'' are 
special cases of (\ref{ffcycl}). For example one has 
\ba
&\nspace\nspace& {}^{\rm in}\!\bra \bar{\th}_n,\bar{a}_n|\, 
\cO \,|\th_{n-1},a_{n-1}; \ldots ;\th_1,a_1 \ket^{\rm out} =  
{}^{\rm out}\!\bra 0|\,\cO\,|\th_n+i\pi ,a_n; \ldots; \th_1,a_1
\ket^{\rm out}\;.  
\nonum
&\nspace\nspace&{}^{\rm out}\!\bra 0|\,\cO\,|
\th_{n-1},a_{n-1};\ldots ;\th_1,a_1;\th_n+2\pi i ,a_n\ket^{\rm out}=
\eta\,{}^{\rm out}\!\bra 0|\,\cO\,|\th_n,a_n; \ldots; \th_1,a_1
\ket^{\rm out}\;.  
\ea
Analogues with `in' and `out' scattering states exchanged follow 
from (\ref{Jscatt}).

The purpose of this paper was to provide a quantum field theoretical 
derivation of the cyclic form factor equation (1.1) or (\ref{ffcycl}).
The derivation given shows that it is a generic feature -- not tied to
integrability -- of massive 1+1 dim.~QFTs with a proper relativistic 
scattering theory. Keeping this in mind, we propose retaining the 
term ``cyclic form factor equation'' for it. The main technical tool 
in the derivation was the use of a family of Rindler spacetimes 
$t \ra R^t$, comoving with the essential support regions 
of the interpolating quantum fields, to transfer the action of geometric 
modular structures to scattering states. We expect that a 4-dim. 
counterpart of the cyclic form factor equation can be derived along 
similar lines, to which we intend to return elsewhere.

\vspace{5mm}
{\tt Acknowledgements:} I wish to thank H.-J. Borchers and 
K.-H. Rehren for discussions, as well as B. Schroer for 
correspondence and a stimulus \cite{Schroer} to write this up. 
In particular I am indebted to K.-H. Rehren for clarifing much of the
material of section 2. The author acknowledges support 
by the Reimar L\"{u}st fellowship of the Max-Planck-Society.


\end{document}